\documentstyle[prl,aps,graphicx]{revtex}

\begin{document}
\title{Adiabaticity in Nonlinear Quantum Dynamics: Bose-Einstein 
Condensate in a Temporally-Varying Box}
\author{Y.\ B.\ Band$^{\,1}$, Boris Malomed$^{\,2}$, Marek 
Trippenbach$^{\,1,3}$}
\address{
$^{\,1}$ Departments of Chemistry and Physics, Ben-Gurion
University of the Negev, 84105 Beer-Sheva, Israel \\
$^{\,2}$ Department of Interdisciplinary Studies, Faculty of
Engineering, Tel Aviv University, Tel Aviv 69978, Israel \\
$^{\,3}$ Institute of Experimental Physics, Optics Division,
Warsaw University, ul.~Ho\.{z}a 69, Warsaw 00-681, Poland
}
\maketitle

\begin{abstract}
A simple model of an atomic Bose-Einstein condensate in a box whose
size varies with time is studied to determine the nature of
adiabaticity in the nonlinear dynamics obtained within the
Gross-Pitaevskii equation (the nonlinear Schr\"{o}dinger equation). 
Analytical and numerical methods are used to determine the nature of
adiabaticity in this nonlinear quantum system.  Criteria for validity
of an adiabatic approximation are formulated.
\end{abstract}


\section{Introduction}

The Adiabatic Theorem of quantum mechanics insures that an eigenstate
of a system whose Hamiltonian evolves sufficiently slowly in time (as
determined by criteria for the applicability of the theorem) will
remain in the same eigenstate, even though the eigenstate evolves in
time \cite {LLQM,Messiah,CohenTan}.  Hence, a slowly evolving system
which is initially in its ground state will remain in the ground state
throughout the course of its evolution.  The adiabatic theorem relies
heavily on the superposition principle of quantum mechanics (although
in classical mechanics similar theorems are valid for nonlinear
systems \cite{LLMech}).  It is of interest to determine to what extent
adiabaticity carries over to {\em nonlinear} quantum systems, such as
Bose-Einstein condensates (BECs) in the region where the mean-field
description is appropriate.  Well below the critical temperature, the
mean-field description is based on the Gross-Pitaevskii equation
(GPE),
\begin{equation}
i\hbar \frac{\partial \Psi }{\partial t}=\left[ -\frac{\hbar
^{2}}{2m}\nabla^{2}+V({\bf r},t)+N_{0}U_{0}|\Psi |^{2}\right] \Psi ,
\label{GP}
\end{equation}
and this approximation often yields excellent results for the system
dynamics, even when the external potential $V$ varies with time.  In
Eq.~(\ref{GP}), $U_{0}=4\pi a_{0}\hbar^{2}/m$ is the atom-atom
interaction strength that is proportional to the $s$-wave scattering
length $a_{0}$, and $m$ is the atomic mass.  The parameter $N_{0}$ in
Eq.~(\ref{GP}) is the total number of atoms, and the wave function
$\Psi $ is subject to the normalization $\int |\Psi ({\bf
r},t)|^{2}d{\bf r}=1$ (the normalization integral is a dynamical
invariant of GPE).  Adiabatic considerations regarding the GPE
dynamics have been applied to cold Bosonic atoms trapped in optical
lattices \cite{Jaksch_99,Anderson,Orzel,Cata}, and the formation of
optical lattice gates for quantum computing from atomic BECs \cite
{Brennen_99}.  However, the applicability of the adiabaticity concept
to BECs does not follow from the above-mentioned Adiabatic Theorem of
quantum mechanics, since the nonlinearity does not allow applicability
of the superposition principle to the GPE. On the other hand,
adiabaticity of nonlinear wave equations, and in particular, of
soliton solutions to such equations, have been extensively studied
(for a review, see Ref.~\cite{KM}).  Nevertheless, BEC dynamical
problems based on the GPE have their own specific features, so that
this case can be different from that studied in the framework of
perturbed soliton solutions to the nonlinear Schr\"{o}dinger equation
(NLSE) in other contexts.  Here, we develop a physically relevant
one-dimensional BEC model which we study in detail by means of
analytical and numerical methods to determine the nature of
adiabaticity in nonlinear quantum systems within mean-field theory.

There are several regimes in which adiabaticity can be experimentally
and theoretically probed for nonlinear systems.  The simplest regime
is one for which the characteristic dynamical time scale (i.e., the
time during which parameters of the Hamiltonian undergo an essential
change), $T$, satisfies conditions
\begin{equation}
\tau_{AD} \ll T \ll \tau_{NL}.  \label{applicability}
\end{equation}
Here, $\tau_{AD}$ is the quantum-mechanical linear adiabatic time
scale determined in terms of the inverse of the difference of the
energy eigenvalues at different values of time, $\tau_{AD} = {\rm
\max}\{\hbar/\left[ \epsilon_{1}(t)-\epsilon_{0}(t)\right] \}$, where
the maximum is taken with respect to a given time interval
\cite{Messiah}, while the nonlinear time scale is $\tau_{NL}={\rm \max
}\{\hbar /\mu (t)\}$, with $\mu (t)$ being the instantaneous chemical
potential \cite{Tripp_2000}.  In this case, the applicability of the
linear Adiabatic Theorem is secured by the first inequality in
(\ref{applicability}), and nonlinearity cannot play a significant role
in the dynamics due to the second inequality.  Therefore, the dynamics
must be adiabatic.  In particular, this regime applies to the NIST
optical-lattice experiments wherein light pulses in the $\mu$s range
are applied to a sodium BEC \cite{NIST}.

A more interesting and more problematic regime is when the dynamical
time scale is the {\em largest} one in the problem, i.e.,
$\tau_{AD},\tau_{NL} \ll T$.  This case applies, e.g., to the BEC
experiments reported in Refs.~\cite{Anderson,Orzel}.  Here, the
nonlinearity plays an essential role in the dynamics, and a relevant
question is whether the dynamics can be adiabatic.  Generally, a
further time scale appears in the multidimensional GPE, viz., the
diffraction time, $\tau_{DF}=2mL_{0}^{2}/\hbar$, where $L_{0}$ is the
length of the system (see below).  In what follows, we show that the
GPE does allow for adiabaticity when $\tau_{AD},\tau_{DF},\tau_{NL}\ll
T$, and we give explicit criteria for the validity of adiabaticity in
the GPE. We do so by means of an analytical estimate of corrections to
the adiabatic approximation, and we present numerical results for the
dynamics of this system.

\section{The dynamical model}

We consider a model based on the 1D GPE, in which the external
potential $U(x)$ is an infinitely deep well,
\begin{equation}
U(x)=\left\{ 
\begin{array}{ll}
0, & 0<x<L(t), \\ 
+\infty , & x<0\;{\rm or}\;x>L(t)\ .
\end{array}
\right.  \label{U(x)}
\end{equation}
The size of the well $L(t)$ slowly varies with time, and we are
interested in determining the behavior of the system in this case, to
determine the applicability of adiabaticity.  The 1D GPE takes the
form
\begin{equation}
i\hbar \psi_{t}=-\frac{\hbar^{2}}{2m}\,\psi_{xx}+g\left| \psi \right|
^{2}\psi \ ,  \label{GPE}
\end{equation}
with the boundary conditions 
\begin{equation}
\psi (0,t)=\psi (L(t),t)=0\ ,  \label{bcL}
\end{equation}
and normalization of the wave function, 
\begin{equation}
\int_{-\infty }^{\infty }|\psi (x)|^{2}dx=1\ .  \label{1}
\end{equation}
The nonlinearity parameter $g$ appearing in this 1D GPE is related to
the nonlinearity parameter $N_{0}U_{0}$ in its 3D counterpart,
Eq.~(\ref{GP}), and is determined so that $N_{0}U_{0}|\Psi_{m}|^{2} =
g|\psi_{m}|^{2}$, where $\Psi_{m}$ and $\psi_{m}$ are the maximum
values of the 3D and 1D wave functions, respectively.  This condition
insures that the time scales for the nonlinear interaction in the 3D
and 1D cases are equal (see below and Ref.~\cite {Tripp_2000}).

This is the generalization of the particle in a box problem to the
case where (a) the size of the box is varying with time, and (b) there
are many bosonic particles in the box that are interacting via a mean
field.

A typical situation in which the dynamics may be adiabatic is when the
function $L(t)$ takes on constant values as $t\rightarrow \pm \infty
$, and slowly varies in between on a long time scale $T$.  We aim to
find the final state $\psi(x,t=+\infty )$ into which an initial state
$\psi(x,t=-\infty )$ will be transformed if $T$ is sufficiently
large, and to check whether the wave function $\psi(x,t)$ remains
adiabatic during the course of the evolution, provided that the function
$L(t)$ varies slowly enough (in practice, of course, the evolution
time interval is large but finite).  To determine what ``sufficiently
slow'' means, we define the nonlinear time scale obtained directly
from the GPE as \cite{Tripp_2000}
\begin{equation}
t_{NL}=(g|\psi_{m}|^{2}/\hbar )^{-1}\approx \mu /\hbar \ ,
\label{tNL}
\end{equation}
where $\mu $ and $|\psi_{m}|$ are the chemical potential and maximum
of the wave function in the initial configuration.  The evolution is
is slow as compared to nonlinear time scale if $T\gg t_{NL}$.  In many 
BEC systems, the nonlinear time scale is large compared with the 
diffraction time scale $\tau_{DF}=2mL_{0}^{2}/\hbar$, also obtained 
directly from the GPE \cite{Tripp_2000}, so we should have $T \gg t_{NL} 
\gg \tau_{DF}$.

It is convenient to transform the variables $x$, $t$ and $\psi $ to
new (dimensionless) variables $\tau $, $\xi $ and $u$:
\begin{equation}
\xi \equiv x/L(t)\ ,  \label{xi}
\end{equation}
\begin{equation}
\tau \equiv \frac{\hbar }{2m}\int_{0}^{t}\frac{dt^{\prime
}}{L^{2}(t^{\prime })}\ , \label{tau}
\end{equation}
\begin{equation}
u\equiv \sqrt{2gm}\hbar^{-1}L(t)\,\psi \ .  \label{u}
\end{equation}
Note that the problem is mapped onto a fixed spatial interval $\xi \in
\lbrack 0,1]$ of the dimensionless spatial variable $\xi $, and the
boundary condition is therefore not time dependent when the problem is
reformulated in terms of these variables.  The 1D GPE~(\ref{GPE})
takes the following form in terms of the new variables:
\begin{equation}
iu_{\tau }+u_{\xi \xi }-\left| u\right|^{2}u = i\left( L_{\tau
}/L\right) \left( \xi u\right)_{\xi }\ , \label{eq_u}
\end{equation}
where $L_{\tau }\equiv dL/d\tau $.  Equation (\ref{eq_u}) is
supplemented by boundary conditions following from Eq.~(\ref{bcL}),
\begin{equation}
u(\xi =0,\tau )=u(\xi =1,\tau )=0\ .  \label{bc1}
\end{equation}

The norm defined in terms of the transformed wave function $u$, 
\begin{equation}
N[u(\tau )]\equiv \int_{0}^{1}\left| u(\xi ,\tau )\right|^{2}d\xi \,,
\label{normN}
\end{equation}
is {\em not} conserved in time, unlike the original norm,
$\int_{0}^{L(t)}\left| \psi (x,t)\right|^{2}dx=1$.  Indeed, as follows
from the substitution of Eq.~(\ref{u}) for $u$ into Eq.~(\ref{normN}),
the $u$-norm is an explicit function of time:
\begin{equation}
N[u(\tau )]=\frac{2gm}{\hbar^{2}}L(\tau )\equiv n_{0}\frac{L(\tau
)}{L_{0}}\ , \label{NN}
\end{equation}
where $L_{0}\equiv L(t=-\infty )$ (or alternatively $L_{0}\equiv
L(t=t_{0})$ if the initial moment in time is $t_{0}$).  The
dimensionless nonlinear-strength parameter,
\begin{equation}
n_{0}\equiv 2gmL_{0}/\hbar^{2},  \label{n0}
\end{equation}
introduced in Eq.~(\ref{NN}) will play an important role below.

When the system size $L(\tau)$ is a slowly varying function of time,
the right-hand side (RHS) of Eq.~(\ref{eq_u}) is small, being
proportional to the logarithmic derivative of the slowly varying
function.  Therefore Eq.~(\ref{eq_u}) may be naturally considered as a
perturbed self-defocusing NLSE, and the adiabatic methods for
nonlinear wave equations reviewed in Ref.~\cite{KM} might be applied. 
However, the perturbation term on the RHS of Eq.~(\ref{eq_u}) need not
allow straightforward application of the perturbation theory to the
present problem since this term does {\em not} vanish at $\xi =0$ and
$\xi =1$ when a general solution found in the zeroth-order
approximation (the expression (\ref{zeroth}) below) is inserted into
it.  One can easily check that, as a consequence of this, a
perturbative expansion generated by the term on RHS of
Eq.~(\ref{eq_u}) is incompatible with the boundary conditions
(\ref{bc1}).

To resolve the problem, we transform the wave function once again,
defining
\begin{equation}
u(\xi ,\tau )\equiv v(\xi ,\tau )\,\,\exp \left(
\frac{i}{4}\frac{L_{\tau }}{L}\xi^{2}\right) .  \label{uv}
\end{equation}
The transformation (\ref{uv}) generates a more convenient form of the
perturbed NLS equation, 
\begin{equation}
iv_{\tau }+v_{\xi \xi }-\left| v\right|^{2}v=i\frac{L_{\tau
}}{2L}v+\frac{LL_{\tau \tau }-2L_{\tau }^{2}}{4L^{2}}\,\xi^{2}v\ ,
\label{v}
\end{equation}
which is subject to the same boundary conditions as in
Eq.~(\ref{bc1}), $v(\xi =0,\tau )=v(\xi =1,\tau )=0$.  An obvious
advantage of having the perturbed NLS equation in the form (\ref{v})
is that now the perturbation vanishes at $\xi =0$ and $\xi =1$, once a
solution found in the zeroth-order approximation vanishes at these
points.

Note that the first term on the right--hand side of Eq.~(\ref{v}) is
non-conservative.  Accordingly, it is straightforward to see that this
term leads to the exact relation (\ref{NN}) for the norm evolution. 
Another important fact is that the second term on the RHS of
Eq.~(\ref{v}), unlike the first term, is {\em second-order small} with
regard to derivatives of the slowly varying functions.  In the
perturbation-theory section that follows below, we will not consider
effects produced by the second-order term, focusing solely on the most
important first-order effects.

\section{Adiabatic Perturbation theory}

\subsection{Zeroth-Order Approximation}

In zeroth-order approximation of the perturbation theory (neglecting
the RHS of Eq.~(\ref{v})), an exact stationary solution satisfying the
zero boundary conditions at $\xi =0$ and $\xi =1$ is given by
\cite{manual,notation,Carr}:
\begin{equation}
v(\xi ,\tau )=2^{3/2}kK(k)\ {\rm sn}(2K(q)\xi ,k)\ \exp (-i\mu \tau
)\equiv V(\xi ;k)\,\exp \left( i\phi \left( \tau \right) \right) \ . 
\label{zeroth}
\end{equation}
Here ${\rm sn}(\cdot ,\cdot )$ is the doubly periodic Jacobi elliptic
sine function, $k$ is the corresponding elliptic modulus, $K(k)$ is the
complete elliptic integral of the first kind, and the chemical
potential $\mu $ is related to $k$ as follows:
\begin{equation}
\mu =4(1+k^{2})K^{2}(k)\ .  \label{mu}
\end{equation}
The modulus $k$, which takes values $0\leq k\leq 1$, determines the
strength of the nonlinearity: it is weak if $k\rightarrow 0$, and
strong if $k\rightarrow 1$.  In fact, $k$ is related directly to the
dimensionless nonlinearity-strength parameter $n_{0}$ defined in
Eq.~(\ref{n0}) as follows:
\begin{equation}
8K(k)\left[ K(k)-E(k)\right] =n_{0}\ ,  \label{k_n0_basic}
\end{equation}
where $E(k)$ is the complete elliptic integral of the second kind. 
Thus, $k$ completely determines the normalization of the initial wave
function, and visa versa.

\begin{figure}[!htb]
\centerline{\includegraphics[width=3in,keepaspectratio]{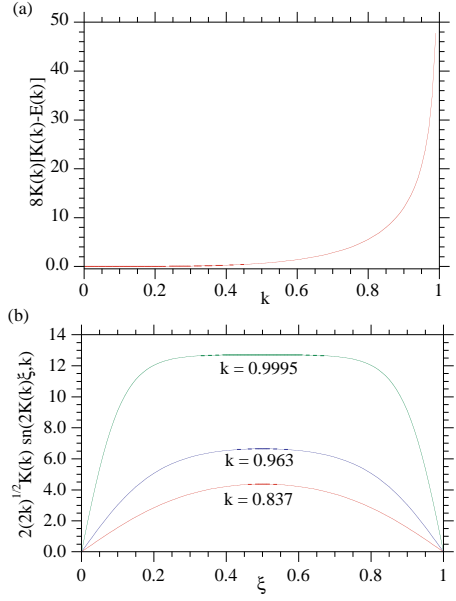}}
\caption{(a) The expression $8K(k)\left[K(k)-E(k)\right]$ versus the
elliptic modulus $k$.  (b) The zeroth-order analytic solution
$2^{3/2}kK(k)\ {\rm sn}(2K(q)\protect\xi ,k)$ for three different
values of $k$.  The normalization of these soliton solutions can be
read off the curve in (a).}
\label{fig1}
\end{figure}

To illustrate the zeroth-order solution, plots of $8K(k)\left[
K(k)-E(k) \right] $ versus $k$ (see Eq.~(\ref{k_n0_basic})), and
$2^{3/2}kK(k)\ {\rm sn} (2K(q)\xi ,k)$ versus $\xi $ for three
different values of $k$ in the regime of strong nonlinearity, $k>0.5$
(see Eq.~(\ref{zeroth})), are displayed in Figs.~\ref{fig1}(a) and
\ref{fig1}(b).  We remark that an exact solution that can be expressed
in terms of the Jacobi elliptic functions is frequently called a {\it
cnoidal wave}, which stems from the notation ${\rm cn}$ for the
Jacobi's elliptic cosine, related to the elliptic sine.

\subsection{The Nonlinear Adiabatic Approximation}

The first-order perturbation term on the RHS of Eq.~(\ref{v}) can be
treated in terms of nonlinear adiabatic perturbation theory \cite{KM}. 
We stress that, unlike the perturbation term in the intermediate
equation (\ref{eq_u}), which is ``abnormal'' in the sense that it is
not compatible with the necessary boundary conditions, as it was
explained above, the ``normal'' perturbation in Eq.~(\ref{v})
satisfies the boundary conditions.  The applicability of simple
perturbative techniques for this class of models can be proved using a
rigorous expansion based on the inverse scattering transform for the
unperturbed NLS equation (i.e., one can prove that the ``simple
techniques'' yield, in the lowest-order nontrivial approximation,
exactly the same results as the rigorous methods, see Ref.~\cite{KM}
and references therein).

The first standard step of the perturbative analysis is to apply the
lowest-order adiabatic approximation.  This approximation takes the
unperturbed solution (\ref{zeroth}), which contains the parameter $k$,
and makes it the first-order approximate solution to the perturbed
equation, assuming that the modulus $k$ is slowly varying in time,
rather than remaining constant.

The slow dependence of the parameter(s) is introduced so as to cancel
the secular divergence(s) in the perturbation theory.  An important
case is when the unperturbed solution contains a single nontrivial
parameter ($k$, in the present case), and the perturbed equation gives
rise to an exact relation replacing a conservation law existing in the
unperturbed version of the equation (this exact relation is usually
called a {\it balance equation} for the (former) conserved quantity). 
This is the case in Eq.~(\ref{NN}).  Then, the time dependence of the
parameter, i.e., $k(\tau )$, can be found in a very simple way by
substitution of the zeroth-order approximation for the solution into
the balance equation \cite{KM}.  In the present case, this condition
amounts to evaluation of the actual value of the norm (\ref{normN}),
inserting the solution (\ref{zeroth}) into it, and then substituting
the result into the exact relation (\ref{NN}).  The final result is
\begin{equation}
8K(k)\left[ K(k)-E(k)\right] =n_{0}\,L(\tau )/L_{0}\ .  \label{basic}
\end{equation}
Eq.~(\ref{basic}) is a transcendental equation to determine $k(\tau)$
for a given function $L(\tau)$ and $n_{0}$ (recall that $n_{0}$ is a
constant).

An essential ingredient of the adiabatic approximation is a consistent
definition of the phase $\phi(\tau)$ for the first-order solution with
variable $k(\tau)$.  Indeed, substituting $k(\tau)$ back into the
general expressions (\ref{zeroth}) and (\ref{mu}) for the wave
function, it is easy to see that the consistently defined phase is
\begin{equation}
\phi (\tau )=-\int_{\tau_{0}}^{\tau }\mu (\tau^{\prime })d\tau^{\prime
}\equiv -4\int_{\tau_{0}}^{\tau }\left[ 1+k^{2}(\tau^{\prime })\right]
K^{2}\left( k(\tau^{\prime })\right) \,d\tau^{\prime }\ ,  \label{phi}
\end{equation}
$\tau_{0}$ being the initial time ($\tau_{0}=-\infty $ in the usual
formulation of the adiabatic approximation).

Thus, the full expression for the lowest-order perturbative solution
obtained in the adiabatic approximation is 
\begin{equation}
v(\xi ,\tau )=V\left( \xi ;k(\tau )\right) \,\exp \left( i\phi (\tau
)\right) \ ,  \label{first}
\end{equation}
where $V(\xi ;k)$ and $\phi(\tau )$ are given by Eqs.~(\ref{zeroth})
and (\ref{phi}), respectively.  Note that expression (\ref{first})
automatically satisfies the zero boundary conditions at the points
$\xi =0$ and $\xi =1$.

Knowing a particular form of the slow temporal dependence $k(\tau)$
obtained from Eq.~(\ref{basic}), one can find the temporal dependence
of the solution's amplitude,
\begin{equation}
A(\tau )\equiv \max_{\xi }|v(\xi ,\tau )|=2^{2/3}k(\tau )K\left(
k(\tau )\right) \ .  \label{A}
\end{equation}
The temporal dependence of state's width (which, for instance, can be
defined as the full width at half-maximum of $|v(\xi ,\tau )|^{2}$)
can similarly be obtained in the adiabatic approximation from the
above expressions.  Using Eqs.~(\ref{mu}) and (\ref{basic}), it is
also possible to predict the evolution of the instantaneous value of
the chemical potential $\mu(\tau)$.

\subsection{Corrections to the Lowest-Order Adiabatic Approximation}

Once the slow time dependence of $k(\tau)$ has been determined as
described above, one can look for perturbation-induced corrections to
the state's shape, which was not taken into account in the first-order
adiabatic approximation.  A solution to Eq.~(\ref{v}) including the
corrections can be sought in the form of an expansion compatible with
the zero boundary conditions, namely,
\begin{equation}
v(\xi ,\tau )=\left[ V(\xi ;k)+\sum_{m=1}^{\infty }b_{m}(\tau )\sin (\pi
m\xi )\right] \exp \left( i\phi (\tau )\right) \ ,  \label{corrections}
\end{equation}
where the functions $V$ and $\phi $ are those which were obtained in the
previous subsection.

The simplest way to derive evolution equations for the amplitudes
$b_{m}(t)$ is to directly substitute the expansion (\ref{corrections})
into Eq.~(\ref{v}), multiply the resulting equation by $\sin(\pi
m\xi)$, and integrate from $\xi =0$ to $\xi =1$, carrying out this
procedure for each integer $m$.  The correction terms are neglected
when substituting the expression (\ref{corrections}) into the first
perturbation term on the RHS of Eq.~(\ref{v}), as they would give rise
to higher-order perturbations.  Implementing this procedure, we use
the classical Fourier expansion for the function ${\rm sn}$,
\begin{eqnarray}
{\rm sn}(2K(k)\xi ,k) &=&\frac{2\pi }{kK(k)}\sum_{p=1}^{\infty
}\frac{Q^{p-1/2}}{1-Q^{2p-1}}\sin \left( \pi (2p-1)\xi \right) ,
\label{Jacobi} \\
Q(k) &\equiv &\exp \left[ -\frac{\pi K\left( \sqrt{1-k^{2}}\right)
}{K(k)} \right] \,.  \label{Q}
\end{eqnarray}
A complicated system of inhomogeneous linear evolution equations for
$b_{m}(\tau )$ ensues.  If $m$ is odd, i.e., $m\equiv 2p-1$, we obtain
\begin{equation}
i\frac{db_{2p-1}}{d\tau }+R_{2p-1}b_{2p-1}-\sum_{n=1}^{\infty
}M_{2p-1,n}\left( 2b_{n}-b_{n}^{\ast }\right) =\frac{iL_{\tau
}}{L}\frac{\pi }{kK\left( k\right) }\frac{Q^{p-1/2}}{1-Q^{2p-1}}\ ,
\label{odd}
\end{equation}
and, if $m$ is even, i.e., $m\equiv 2p$, 
\begin{equation}
i\frac{db_{2p}}{d\tau }+R_{2p}b_{2p}-\sum_{n=1}^{\infty
}M_{2p,n}\left( 2b_{n}-b_{n}^{\ast }\right) =0\ . \label{even}
\end{equation}
Here $Q$ is the {\it Jacobi parameter} defined in Eq.~(\ref{Q}), and
the coefficients appearing on the left-hand sides of Eqs.~(\ref{odd})
and (\ref{even}) are
\begin{eqnarray}
R_{m} &\equiv &2\left( 1+k^{2}\right) K^{2}(k)-\frac{1}{2}\left( \pi
m\right)^{2},  \label{R} \\
M_{mn} &\equiv &4k^{2}K^{2}(k)\int_{0}^{1}{\rm sn}^{2}(2K(k)\xi
,k)\left[ \cos \left( \pi (m-n)\right) -\cos \left( \pi (m+n)\right)
\right] d\xi \ .
\label{M}
\end{eqnarray}
In fact, all the coefficients $M_{mn}$ with $m$ and $n$ having
opposite parities are zero, hence we may set $b_{2p}\equiv 0$, and we
are left with the system of equations (\ref{odd}).  Recall that one
should substitute the time-dependent modulus $k(\tau )$ as found from
Eqs.~(\ref{basic})) into the above expressions.

This cumbersome system can be simplified if $k(\tau )$ does not take
on values too close to unity.  Then, ${\rm sn}$ remains close to the
usual sine function (for instance, at $k^{2}=1/2$, the Jacobi
parameter, which determines the anharmonicity of the expansion
(\ref{Jacobi}), is $Q=\exp (-\pi )\approx 0.043$, which may be
regarded as a sufficiently small expansion factor).  Thus, to obtain a
simple approximation for the coefficients $M_{mn}$ defined in
Eq.~(\ref{M}), one may simply set ${\rm sn} (2K(k)\xi ,k)\approx \sin
(\pi \xi )$.  Within this approximation the only nonzero components of 
in the matrix $\left( M_{mn}\right) $ are 
\begin{equation}
M_{11}=3k^{2}K^{2}(k),\,M_{mm}=2k^{2}K^{2}(k)\,\
(m>1),\,M_{m,m-2}=M_{m-2,m}=-k^{2}K^{2}(k) \ . \label{nonzero}
\end{equation}
Furthermore, the RHS of Eq.~(\ref{odd}) also greatly simplifies in the
same approximation.  It is different from zero solely for $m=1$, being
equal to $iL_{\tau }/(2L)$.  Thus, the approximation which replaces
the ${\rm sn}$ function by the usual sine leads to the following
equations, instead of Eqs.~(\ref{odd}) and (\ref{even}):
\begin{equation}
i\frac{db_{1}}{d\tau }+\left[ 2\left( 1-2k^{2}\right)
K^{2}(k)-\frac{\pi^{2} }{2}\right] b_{1}+3k^{2}K^{2}(k)b_{1}^{\ast
}+k^{2}K^{2}\left( 2b_{3}-b_{3}^{\ast }\right) =\frac{iL_{\tau
}}{2L}\,, \label{p=1}
\end{equation}
\[
i\frac{db_{2p-1}}{d\tau }+\left[ 2\left( 1-k^{2}\right) K^{2}(k)
-\frac{[\pi (2p-1)]^{2}}{2}\right]
b_{2p-1}+2k^{2}K^{2}(k)b_{2p-1}^{\ast }
\]
\begin{equation}
+k^{2}K^{2}\left( 2b_{2p-3}+2b_{2p+1}-b_{2p-3}^{\ast }-b_{2p+1}^{\ast
}\right) =0\,,  \label{2p-1}
\end{equation}
where $p>1$.  Recall that all the amplitudes $b_{m}$ with even values
of $m$ are zero.

Despite the fact that the approximate system consisting of Eqs.~(\ref
{p=1}) and (\ref{2p-1}) is considerably simpler than the exact
Eqs.~(\ref{odd}), it can only be solved numerically by truncating the
system of the linear equations at some finite integer.  Nevertheless,
some qualitative generic features of the solution can be determined. 
The general structure of the system is of the form
\begin{equation}
i\frac{dB}{d\tau }+{\bf A}(\tau )B = iC(\tau )\ ,  \label{general}
\end{equation}
where $B$ is a column vector of the variables $b_{2p-1}$, ${\bf A}$ is
a matrix of coefficients multiplying the variables $b$, and $C$ is the
vector column of free terms on the left-hand side, with the single
nonzero entry $c_{1}\equiv L_{\tau }/(2L)$.  Both $C$ and ${\bf A}$
slowly depend upon time - the former directly, the latter via $k(\tau
)$.

Solutions to the system (\ref{general}) consist of terms of the type 
\begin{equation}
\int_{\tau_{0}}^{\tau }\exp \left( -i\int^{\tau^{\prime }}\omega (\tau
^{\prime \prime })d\tau^{\prime \prime }\right) \,f(\tau^{\prime })d\tau
^{\prime }\ ,  \label{integrals}
\end{equation}
where $\omega (\tau)$ is an eigenvalue of the matrix ${\bf A}(\tau)$,
and $f(\tau)$ are slowly varying functions similar to the
above-mentioned $c_{1}$.  Note that the time scales $2\pi /\omega $,
determined by different eigenfrequencies $\omega $, are, in fact, a
mixture of the adiabatic and nonlinear time scales, $\tau_{AD}$ and
$\tau_{NL}$, defined in the Introduction.

The following conclusion can be made concerning the size of the {\em
nonadiabatic} effects (shape corrections) considered above.  If the
function $L(\tau )$ slowly depends on $\tau $ with a characteristic
time scale $T$ (as defined in the introduction), and if a
characteristic value of $\omega$ is $\omega_{0}$ (within the limits
of its slow evolution on the time scale $\ \sim T$), the criterion for
the applicability of the adiabatic approximation is
\begin{equation}
\omega_{0}T/(2\pi )\gg 1\ .  \label{condition}
\end{equation}
We stress that, as the characteristic times $2\pi /\omega_{0}$ taken
for the different eigenfrequencies constitute a set including the
adiabatic and nonlinear time scales $\tau_{AD}$ and $\tau_{NL}$ (see
above), the inequality (\ref{condition}) is exactly the condition for
the applicability of the adiabatic approximation conjectured in the
Introduction.

The evolution equations (\ref{general}) for the shape-correction
amplitudes are to be solved for an initial state without shape
corrections, i.e., $b_{m}(\tau =\tau_{0})=0$ $\forall \ m$.  If one
takes the initial moment as $\tau_{0}\rightarrow -\infty $ (as
mentioned above, this is the standard assumption in the treatment of
adiabatic processes \cite{LLQM,LLMech}), one can determine eventual
values of the shape-correction amplitudes as $b_{m}(\tau )$ at $\tau
\rightarrow +\infty $.  Classical estimates for integrals involving
products of rapidly and slowly varying functions \cite {LLQM} show
that the values of $b_{m}(\tau \rightarrow +\infty )$ are {\em
exponentially small} when condition (\ref{condition}) is satisfied:
\begin{equation}
|b_{m}(\tau =+\infty )|\sim \exp \left( -{\rm const}\cdot \omega
_{0}T\right) \ .  \label{estimate}
\end{equation}
A particular value of the constant in this expression depends on the
choice of the unperturbed state and on the form of the function
$L(\tau )$.  Hence, in analogy with the well-known theorems estimating
nonadiabatic corrections to the adiabatic approximation in (nonlinear)
classical mechanics \cite{LLMech}, the $b_{m}(\tau \rightarrow +\infty
)$ values are exponentially small.

\section{Numerical Results}

\begin{figure}[tbp]
\centerline{\includegraphics[width=3in,angle=270,keepaspectratio]{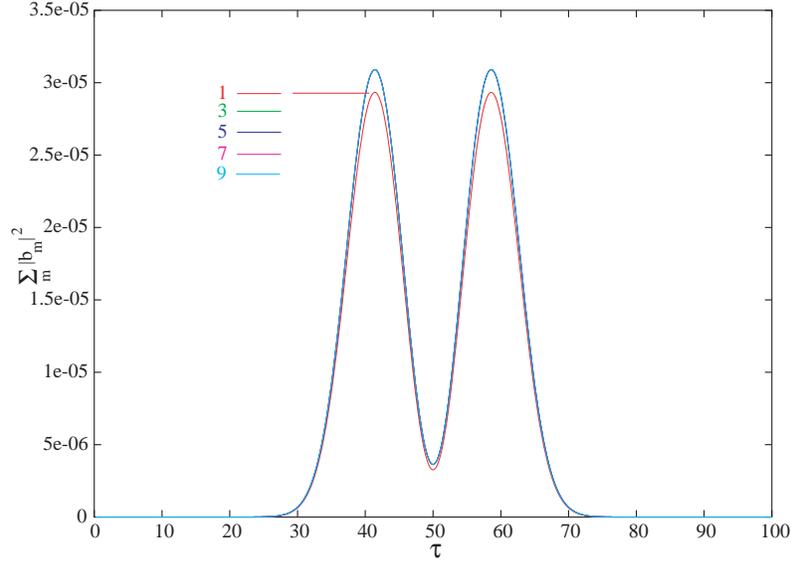}}
\caption{The probability $\sum_{m=1}^{N_c} |b_m(\protect\tau)|^2$ for
the nonadiabatic correction to the state versus $\protect\tau$ for
$N_c = 1,3,5,7,9$ with $T = 100$.  For $N_{c} \ge 3 $ the curves lie
on top of each other.}
\label{fig2}
\end{figure}

\begin{figure}[tbp]
\centerline{\includegraphics[width=3in,angle=270,keepaspectratio]{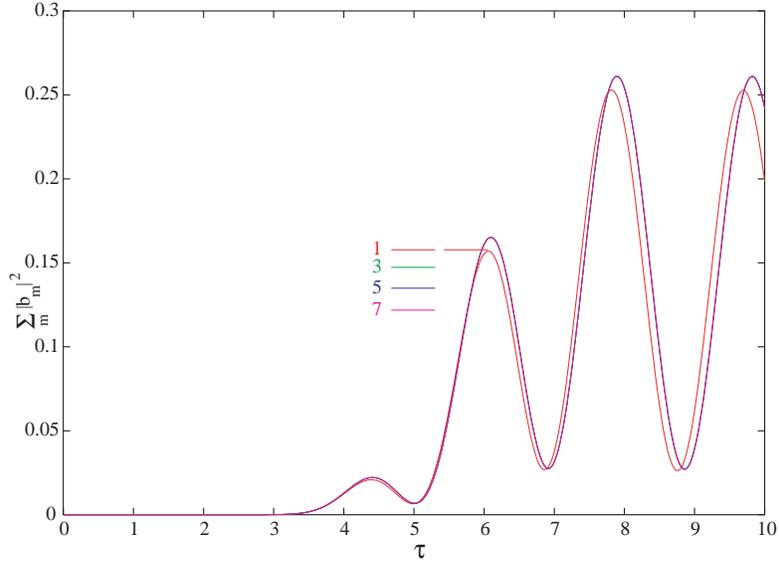}}
\caption{Same as Fig.~\ref{fig2} except for $T = 10$.}
\label{fig3}
\end{figure}

We first present results for the amplitudes $b_{m}(\tau)$ in Eq.~(\ref
{corrections}), obtained by numerically solving Eqs.~(\ref{p=1}) and
(\ref {2p-1}).  We take $k^{2}=0.3$, $L_{0}=1$, and
\[
L(\tau ) = L_{0}\left[ 1 + \exp \left( -\frac{(\tau
-T/2)^{2}}{2\sigma^{2}}\right) \right] \,,
\]
with $\sigma = T/10$.  Figure~\ref{fig2} shows the computed
excited-state probability,
\begin{equation}
P_{{\rm ex}}(\tau ) \equiv \sum_{m=1}^{N_{c}}|b_{m}(\tau )|^{2},
\label{excited}
\end{equation}
versus time $\tau $, with the number of modes kept in the truncated
calculation being $N_{c}=1,3,5,7$, and $9$, for $T=100$.  Except for
$N_{c}=1$, all the curves lie on top of each other, hence the results
do converge very quickly as a function of the number of the modes.  We
see from Fig.~\ref{fig2} that the probability of finding excited
states for all times is below $3.2\times 10^{-5}$, and for $t=T$ the
probability is exceedingly small, i.e., the process is almost
completely adiabatic.  A minimum of $P_{{\rm ex}}(\tau =T/2)$ is
expected from the general form of the perturbation equations since the
derivative of $L(\tau )$ vanishes at $\tau = T/2$.  For $T\,\,_{\sim
}^{<}\,10$, the excited-state probability (\ref{excited}) begins to be
large ($>0.2$), and the adiabatic-theory results are no more reliable. 
For example, Fig.~\ref{fig3} shows the results for $T = 10$.  Again
the convergence as a function of the number of the modes is very fast,
but the excited state probability is not small.  For times $\tau > T$,
$P_{{\rm ex}}(\tau)$ oscillates with time.

\begin{figure}[tbp]
\centerline{\includegraphics[width=3in,angle=270,keepaspectratio]{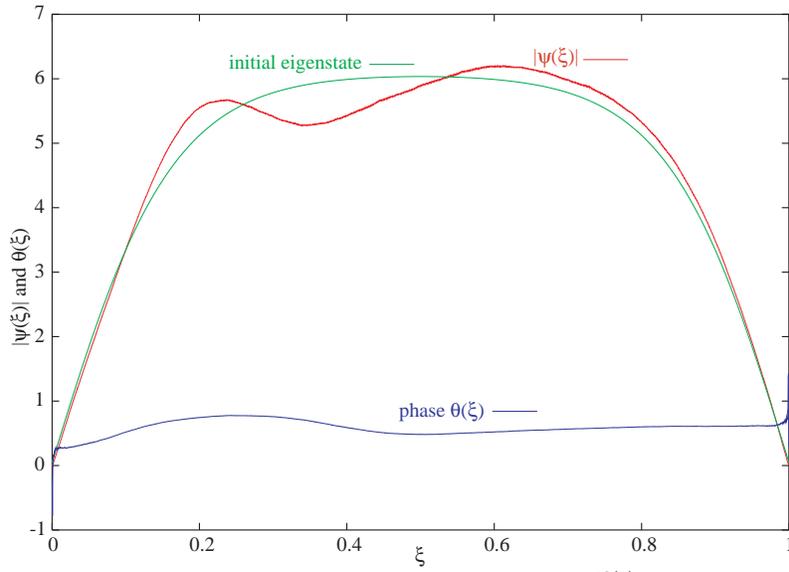}}
\caption{The magnitude and phase of the wave function
$\protect\psi(\protect\xi, \protect\tau=T) =
|\protect\psi(\protect\xi)| e^{i\protect\theta(\protect\xi )}$ versus
the coordinate $\protect\xi$ in the box at the completion of the
dynamical process, with $k=0.963$ and $T=1\times 10^{-2}$.  Also shown
is the magnitude in the initial eigenstate.}
\label{fig4}
\end{figure}

\begin{figure}[tbp]
\centerline{\includegraphics[width=3in,angle=270,keepaspectratio]{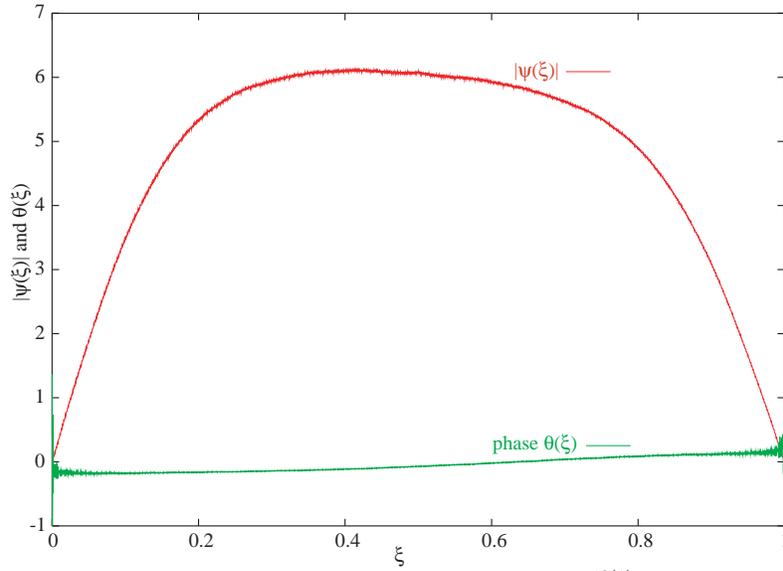}}
\caption{The magnitude and phase of the wave function
$\protect\psi(\protect\xi, \protect\tau=T) =
|\protect\psi(\protect\xi)| e^{i\protect\theta(\protect\xi )}$ versus
the coordinate $\protect\xi$ in the box at the completion of the the
dynamical process, with $k=0.963$ and $T=1$.}
\label{fig5}
\end{figure}

For stronger nonlinearity, $0.5<k^{2}\leq 1$, perturbative methods
cannot be used (in particular, the approximation based on the
replacement of the elliptic sine by the ordinary sine, as described in
the above section, does not apply), so we directly solved
Eq.~(\ref{v}) using a split-step fast Fourier transform method, in
order to check if adiabaticity still takes place in this regime. 
Figure \ref{fig4} shows the results for the calculated wave function
$\psi(\xi ,\tau =T)\equiv |\psi(\xi)|e^{i\theta(\xi )}$ versus $\xi $
in the box at the completion of the dynamical process for small
(non-adiabatic) time-scale, $T=0.01$, and for $k=0.963$.  Also shown
for comparison is the initial eigenstate magnitude
$|\psi(\xi,\tau=0)|$.  The magnitude of the wave function in the final
state is not too different from that in the initial eigenstate, and
the spatial variation of the phase is fairly flat.  Figure~\ref{fig5}
pertains to the same case, but with a larger time scale, $T=1$.  Now,
the magnitudes of the wave function in the finite and initial state
are barely distinguishable on the scale of the figure, and the spatial
profile of the phase is almost flat.  Thus, the process is largely
adiabatic in the latter case.

\section{Conclusion}

We have presented a consistent derivation of the nonlinear dynamics
for a simple model describing a BEC confined in a box with a
temporally-varying size $L(\tau )$.  The speed of the variation of
$L(\tau )$ determines whether the dynamics is adiabatic.  A
``trivial'' regime of adiabaticity is that for which $\tau_{AD}\ll
T\ll \tau_{NL}$; in this work, we have shown that adiabaticity can
also be maintained when $\tau_{AD},\tau_{DF},\tau_{NL} \ll T$, where
the various time scales have been defined in the Introduction.  If
other time scales appear, the condition $\tau_{AD},\tau_{DF},
\tau_{NL}\ll T$ may not be sufficient to insure adiabaticity.  For
example, if a barrier is present in the middle of the box - e.g., a
repulsive delta-function at $x = L_{0}/2$ - then another time scale,
corresponding to the time of tunneling under the barrier, $\tau_{T}$,
is present in the problem.  If this time scale is long compared with
$\tau_{NL}$ , adiabaticity will not be maintained, and a non-vanishing
spatially-varying phase will develop across the condensate wave
function \cite{Band_01}.  Hence, the issue of adiabaticity in
nonlinear problems must be investigated carefully; the perturbative
techniques reviewed in Ref.~\cite{KM} may be applicable, but
additional considerations may play a role.

The particle in a box is a paradigm problem in one-body quantum
mechanics.  We have extended it to the many-body regime at least
within a mean-field approach, and studied the adiabaticity for such a
system when the size of the box varies with time.  Specifically, we
formulated the criteria for the validity of the adiabatic
approximation for a BEC in a box whose size varies with time,
developed the analytical and numerical tools for investigating
adiabaticity in the dynamics within quantum mean-field theory, and
presented results of calculations for this system.

\acknowledgements
This work was supported in part by grants from the US-Israel Binational
Science Foundation, the Israel Science Foundation, the James Franck
Binational German-Israeli Program in Laser-Matter Interaction, and the
Polish KBN 62/P03/2000/18.

\end{document}